\documentclass[12pt, epsfig]{article}
\usepackage{amsmath,amsfonts,amssymb,amsthm,amstext,amscd,eucal,graphicx , xcolor}
\usepackage{slashbox}
\usepackage[all]{xy}
\usepackage{dsfont}
\usepackage{hyperref}
\usepackage{amsmath}
\usepackage{slashed}
\makeatletter \@addtoreset{equation}{section}

\makeatletter\renewcommand\section{\@startsection {section}{1}{\z@}%
                                   {-3.5ex \@plus -1ex \@minus -.2ex}
                                   {2.3ex \@plus.2ex}%
                                   {\normalfont\large\bfseries}}
\renewcommand\subsection{\@startsection{subsection}{2}{\z@}%
                                     {-3.25ex\@plus -1ex \@minus -.2ex}%
                                     {1.5ex \@plus .2ex}%
                                     {\normalfont\bfseries}}

\newcommand{\be}{\begin{equation}}
\newcommand{\ee}{\end{equation}}
\newcommand{\bea}{\begin{eqnarray}}
\newcommand{\eea}{\end{eqnarray}}
\newcommand{\bse}{\begin{subequations}}
\newcommand{\ese}{\end{subequations}}
\newcommand{\beqa}{\begin{eqnarray}}
\newcommand{\eeqa}{\end{eqnarray}}
\newcommand{\beqar}{\begin{eqnarray*}}
\newcommand{\eeqar}{\end{eqnarray*}}
\newcommand{\bi}{\begin{itemize}}
\newcommand{\ei}{\end{itemize}}
\newcommand{\bn}{\begin{enumerate}}
\newcommand{\en}{\end{enumerate}}
\newcommand{\fixme}[1]{\textbf{FIXME: }$\langle$\textit{#1}$\rangle$}

\newcommand{\ba}{\begin{array}}
\newcommand{\ea}{\end{array}}
\newcommand{\bc}{\begin{center}}
\newcommand{\ec}{\end{center}}

\newcommand{\bQ}{\boldsymbol{Q}}
\newcommand{\bT}{\boldsymbol{T}}

\newcommand{\cJ}{\mathcal J}
\newcommand{\cN}{\mathcal N}
\newcommand{\bcJ}{\boldsymbol{\mathcal J}}
\newcommand{\bcL}{\boldsymbol{\mathcal L}}
\newcommand{\bJ}{\boldsymbol{J}}
\newcommand{\lbJ}{\mbox{\large{$\boldsymbol{J}$}}}
\newcommand{\bL}{\boldsymbol{L}}

\newcommand{\cB}{\mathcal B}

\newcommand{\cF}{\mathcal F}
\newcommand{\cH}{\mathcal H}

\newcommand{\HNH}{\mathcal H_{\cN\cH}}

\newcommand{\cL}{{\mathcal{L}}}

\definecolor{darkgreen}{rgb}{0,0.3,0}
\definecolor{darkblue}{rgb}{0,0,0.3}
\definecolor{darkred}{rgb}{0.7,0,0}

\newcommand{\new}[1]{{#1}}
\newcommand{\old}[1]{}

\begin{document}

\begin{titlepage}

\begin{flushright}\vspace{-3cm}
{
IPM/P-2016/nnn  \\
\today }\end{flushright}
\vspace{-.5cm}

\begin{center}
\Large{\bf{\hspace*{-.5cm}Horizon Fluffs:}}\\
\medskip
\large{\bf{Near Horizon Soft Hairs as Microstates of Generic AdS$_3$ Black Holes}}

\bigskip\bigskip

\large{\bf{M.M. Sheikh-Jabbari\footnote{e-mail:
jabbari@theory.ipm.ac.ir}$^{;\ a}$
and  H. Yavartanoo\footnote{e-mail: yavar@itp.ac.cn }$^{;\ b}$  }}
\\

\vspace{5mm}
\normalsize
\bigskip\medskip
{$^a$ \it School of Physics, Institute for Research in Fundamental
Sciences (IPM),\\ P.O.Box 19395-5531, Tehran, Iran}
\smallskip

{$^b$ \it  State Key Laboratory of Theoretical Physics, Institute of Theoretical Physics,
Chinese Academy of Sciences, Beijing 100190, China.
 }\\
\date{\today}

\end{center}
\setcounter{footnote}{0}

\begin{abstract}

\noindent
In \cite{Afshar:2016uax} the  \emph{horizon fluffs} proposal is put forward to identify microstates of {generic} non-extremal three-dimensional Ba\~nados--Teitelboim--Zanelli (BTZ) black holes. The proposal is that black hole microstates, which were dubbed as horizon fluffs, are states  labelled by the conserved charges associated with non-trivial diffeomorphisms on the near horizon geometry and these states are not in the  coadjoint {orbits} of the asymptotic Virasoro algebra at the Brown-Henneaux central charge, associated with the BTZ black holes.
It is also known that AdS$_3$ Einstein gravity has more general black hole solutions than the BTZ family which are generically described by two periodic, but otherwise arbitrary, holomorphic and anti-holomorphic functions.  We show that these general AdS$_3$ black holes which are typically conformal descendants of the BTZ black holes and are characterised by the associated Virasoro coadjoint orbits, appear as coherent states in the asymptotic symmetry algebra {corresponding to the black hole family}. We apply the horizon fluffs proposal to these generic AdS$_3$ black holes and identify the corresponding microstates. We then perform  microstate counting and compute the entropy. The entropy appears to be an orbit invariant quantity, providing an important check for the horizon fluffs proposal.


\end{abstract}


\end{titlepage}
\renewcommand{\baselinestretch}{1.1}  
\tableofcontents

 \section{Introduction} Thermodynamic behaviour of black holes plus requirement of unitary evolution for black hole formation and evaporation, hint to presence of an underlying black hole microstate  system. 
Black hole microstates are often attributed to a quantum theory of gravity, and to states which have no counterpart in the classical gravity theory. In particular and within string theory, following \cite{Strominger:1996sh}, there have been many efforts to identify certain supersymmetric black hole microstates, e.g., see \cite{Sen:2007qy}. The ideas from string theory seems not to be applicable to the simplest, realistic black holes like Schwarzschild or Kerr.

On the other hand, after the seminal works Bondi and Mentzer, and Sachs (BMS) \cite{BMS}, and Brown and Henneaux \cite{Brown-Henneaux}, we have learned that in any theory with local gauge symmetry, one may be able to define non-local conserved charges { given through surface integrals.} For example see \cite{Lee-Wald, Barnich-Brandt, Thesis-Seraj-Hajian} and references therein. These charges are associated to a certain (measure-zero) subset of local gauge transformations which remain undetermined even after a complete gauge fixing and may hence be dubbed as residual gauge symmetries \cite{Sheikh-Jabbari:2016lzm}. These charges  satisfy an algebra which is generically infinite dimensional and may admit central extension. It is then natural to assume that states in the quantum Hilbert space of theories with local gauge symmetries can also carry these non-local (residual gauge symmetry) charges.
Therefore, there is an extended Hilbert space   which is a direct product of the usual gauge theory Hilbert space (e.g., containing transverse photon states in the case of QED or transverse traceless gravitons in the case of Einstein general relativity) and the states carrying additional quantum numbers associated with residual gauge symmetries \cite{My-seminar}. The states which only differ in their residual gauge symmetry charges, as these charges are associated with gauge transformations, have the same energy and momentum. These charges may, hence,  be also called ``soft charges'' \cite{HPS}. 

In the context of general relativity we are dealing with diffeomorphisms as local gauge symmetries and the soft charges may be associated with the residual diffeomorphisms \cite{Sheikh-Jabbari:2016lzm}. The residual diffeomorphisms are usually identified through prescribing  appropriate fall off behaviour for the ``asymptotic'' field (metric) perturbations caused by the diffeomorphisms. The BMS \cite{BMS} and Brown-Henneaux \cite{Brown-Henneaux}  have shown the prime examples of such analyses. If one can identify the charges associated with residual diffeomorphisms and their symmetry algebra, then geometries which are diffeomorphic to each other but have different residual diffeomorphism charges become physically distinguishable. One may hence explore the idea whether existence of  these soft charges can remedy the pressing questions in the context of black hole physics, like black hole microstate problem or information paradox \cite{HPS}. In this context one may hence call such states as ``soft hair'' \cite{HPS}, inspired by (and in contrast to) the ``(no) hairs'' statements of black hole \cite{no-hair}. 

Motivated by these ideas ``asymptotic symmetry group'' of many different background geometries have been studied, e.g., see \cite{Kerr/CFT, BMS-algebra},  and the corresponding representations, which in this context is more appropriately called (coadjoint) orbits of the algebra, have been analysed \cite{BMS-Orbits, Garbarz}. In the three dimensional (3d) asymptotic AdS$_3$ cases, this symmetry algebra consists of two copies of Virasoro algebra at the Brown-Henneaux central charge \cite{Brown-Henneaux}, whereas for 3d or 4d asymptotic flat geometries  we are dealing with BMS$_3$ and BMS$_4$ algebras \cite{BMS-algebra}. In this work we will focus on asymptotic locally AdS$_3$ geometries with flat 2d cylinder as the 2d boundary metric.

In black hole geometries we have a horizon and, near horizon region of a generic black hole geometry with bifurcate horizon has a 2d Rindler wedge. It has been noted that at the near horizon geometry one has the possibility of imposing non-trivial boundary conditions on the metric fluctuations leading to a near horizon symmetry algebra \cite{Carlip-NH-entropy, NH-algebra-1, NH-algebra-2, NH-algebra-3, Afshar:2016wfy}.\footnote{There seems to be some different choices for the near horizon boundary conditions and hence different near horizon algebras.} This symmetry algebra is generically different than the asymptotic symmetry algebra.\footnote{For the case of degenerate horizon of extremal black holes, instead of the Rindler wedge we get an AdS$_2$ geometry. The near horizon symmetry algebra in this case is different than the generic non-extreme black holes \cite{Lucietti}. } Despite the possibility of different boundary conditions, analysis of \cite{Afshar:2016wfy} suggests that for all different near horizon geometries there are always boundary conditions which lead to  infinite number of copies of Heisenberg algebra. For the case of 3d geometries this is the algebra of creation-annihilation operators of a 2d free boson theory \cite{Afshar:2016wfy}.

3d Einstein gravity has been viewed as testbed for asking questions about quantum gravity and back holes, e.g., see \cite{Carlip:2004ba}. In particular,  after the seminal work of Ba\~nados-Zanelli-Teitelboim (BTZ) \cite{BTZ}, it is known that  3d Einstein gravity with negative cosmological constant (AdS$_3$ gravity) admits black hole solutions. In this work we will focus on AdS$_3$ gravity and consider a more general class of black holes (than the BTZ solution), those within the class of  Ba\~nados geometries \cite{Banados} and analyse their near horizon and asymptotic symmetries, and identify their microstates.

In \cite{Afshar:2016uax} a fairly simple proposal was made for identifying black hole microstates: There are soft hairs labelled by the near horizon symmetry algebra charges. 
There is a certain class of near horizon soft hairs which are not distinguishable by the asymptotic symmetry algebra charges   {associated with the non-extremal black hole solution or its conformal descendants}. The proposal is that, these states which were called \emph{horizon fluffs}, are the black hole microstates. In this work we extend and generalise the analysis of \cite{Afshar:2016uax} for   {non-extremal} black holes in the class of Ba\~nados geometries. This analysis, besides clarifying some aspects,  provides a further non-trivial check of the black hole horizon fluffs proposal: as was discussed in some detail in \cite{Sheikh-Jabbari:2014nya, Sheikh-Jabbari:2016unm, Compere:2015knw} and as we will briefly review in the next section,  Ba\~nados geometries may be classified by the coadjoint orbits  \cite{Vir-Orbits, Balog} of  asymptotic Virasoro symmetry algebra. Then, as is discussed in \cite{Sheikh-Jabbari:2016unm} entropy should be an \emph{orbit invariant} quantity. Identifying microstates for generic black holes in the Ba\~nados family of solutions, we explicitly construct the horizon fluffs and confirm that the entropy is an orbit invariant.

This  paper is organised as follows. In section \ref{section-Banados-review}, we review Ba\~nados geometries and simple facts about them. This is essentially a review of \cite{Sheikh-Jabbari:2016unm, Compere:2015knw}. In section \ref{section-two-algebras}, {we present the   {near horizon algebra associated with generic AdS$_3$ black holes and its relation to the Brown-Henneaux asymptotic symmetry algebra}.
In section \ref{section-Hilbert-spaces}, we construct the Hilbert space of   {the near horizon algebra}. We show that the Ba\~nados geometry is a coherent state in the asymptotic Hilbert space. We also review how the states in asymptotic and near horizon Hilbert spaces fall into  the Virasoro coadjoint orbits. In section \ref{section-BH-microstates}, we present the definition of black hole microstates through an equation over the near horizon Hilbert space. Solving this equation we identify the microstates and perform their counting. Section \ref{section-discussion} summarises our analysis and results and presents an outlook.

\section{{A quick review on Ba\~nados geometries}}\label{section-Banados-review}

Recalling the Lagrangian and equations of motion of AdS$_3$ Einstein gravity,
\be\label{AdS3-gravity-theory}
{\cal L}=\frac{1}{16\pi G} (R+\frac{2}{\ell^2}),\qquad R_{\mu\nu}=-\frac{2}{\ell^2} g_{\mu\nu},
\ee
and that in 3d Ricci tensor completely determines Riemann curvature, the set of solutions to this theory are all locally AdS$_3$ geometries. These solutions are all locally diffeomorphic to each other and can become distinct solutions only through their residual diffeomorphism charges (if they exist). As is now established, thanks to the work of Brown and Henneaux \cite{Brown-Henneaux}, the set of distinct geometries may be fully specified by the boundary behaviour of the geometries. The set of such geometries which obey the standard Brown-Henneaux boundary conditions \cite{Brown-Henneaux} are given as
\cite{Banados}\footnote{There are more relaxed boundary conditions than the Brown-Henneaux ones, leading to geometries with more independent functions, e.g., see \cite{Cedric, Daniel}.} 
\be\label{generic-Banados-geometry}
ds^2=\ell^2 \frac{dr^2}{r^2}- \Big(rdx^+- \frac{\ell^2 L_-(x^-)}{r} dx^-\Big)\Big(rdx^-- \frac{\ell^2 L_+(x^+)}{r} dx^+\Big)
\ee
where  $L_\pm(x^\pm)$ are two arbitrary periodic functions $L_\pm (x^\pm+2\pi)=L_\pm(x^\pm)$. We will call the metrics in \eqref{generic-Banados-geometry} Ba\~nados geometries. The conformal/causal boundary of these geometries is a cylinder parametrised by $t,\phi$, where
$$
x^\pm=t\pm \phi,\qquad \phi\in[0,2\pi].
$$
Note that as discussed in \cite{Sheikh-Jabbari:2016unm} range of $r$ coordinate may be extended to regions with negative $r^2$. 

The special case of constant, positive $L_\pm$ constitute the BTZ black hole \cite{BTZ} family, while constant $-1/4< L_\pm<0$ family are the conic spaces and $L_\pm=-1/4$ is the global AdS$_3$. 

For the generic functions $L_\pm(x^\pm)$,  being locally AdS$_3$, Ba\~nados geometries have six Killing vector fields forming an $sl(2,R)\times sl(2,R)$ algebra. These six Killing vector fields are of the form \cite{Sheikh-Jabbari:2014nya},
\begin{subequations}\label{Killings}
\begin{align}
\chi[K_+,K_-]&=\chi^r \partial_r+\chi^+\partial_++\chi^-\partial_- ,\\
\chi^r=-\frac{r}{2} (K_+'+K_-'  ),\quad
\chi^+=K_+ +& \frac{\ell^2r^2K_-''+\ell^4L_-K_+''}{2(r^4-\ell^4L_+L_-)},\quad \chi^-=K_- + \frac{\ell^2r^2K_+''+\ell^4L_+K_-''}{2(r^4-\ell^4L_+L_-)},
\end{align}
\end{subequations}
where primes denote derivative w.r.t. the argument and $K_\pm$ are two functions obeying the third order differential equation
\be\label{Kpm}
K_\pm''' -4K_\pm' L_\pm-2K_\pm L'_\pm=0.
\ee
One may show that the solutions to the above equation can be specified by the solutions to the second order equations
\be\label{Hills}
\psi_\pm''-L_\pm\psi_\pm=0.
\ee
The above equation with periodic functions $L_\pm$ is called Hill's equation \cite{Hill-Eq} and if we denote the two linearly independent solutions to the Hill's equation by $\psi_{\alpha\pm}, \alpha=1,2$ the three solutions to each equation in \eqref{Kpm} are
\be
K^0_\pm=\psi_{1\pm}\psi_{2\pm},\quad K^1_\pm=\psi_{1\pm}\psi_{1\pm},\quad K^{-1}_\pm=\psi_{2\pm}\psi_{2\pm}.
\ee
Out of these six $K_\pm$'s, recalling the Floquet theorem \cite{Hill-Eq}, the $K^0_\pm$ are always periodic and lead to two globally defined $U(1)$  Killing vector fields \cite{Sheikh-Jabbari:2014nya}. One may associate Komar-type conserved charges to the two Killing vectors \cite{Hajian:2015xlp, Sheikh-Jabbari:2016unm}. 

Within the Ba\~nados family, geometries for which  $\frac{\psi'_{\alpha \sigma}}{\psi_{\alpha\sigma}}$ ($\sigma=\pm,\ \alpha=1,2)$ are real-valued correspond to black holes geometries with event and Killing horizons\footnote{ {Note that although one can always write two independent solutions of the Hill equation as real functions, however here by $\psi_{\alpha\sigma}$ we are referring to a particular form of solutions appearing in the Floquet theorem: $\psi_1(x)=e^{{\cal T} x}P_1(x),\ \psi_2(x)=e^{-{\cal T} x}P_2(x)$  where $P_1, P_2$ are two periodic smooth functions and ${\cal T}$ is the Floquet index (e.g. see \cite{Sheikh-Jabbari:2016unm}). The Flouqet index ${\cal T}$ can be real or pure imaginary. The black hole family we will be discussing here corresponds to real Floquet index case.\label{Footnote-Floquet}}}. BTZ family (with constant $L_\pm$) are very special cases in this class. The four branches of the bifurcate inner and outer Killing horizons 
are located at \cite{Sheikh-Jabbari:2016unm}
\be\label{horizon-radii}
r_{\alpha\beta}^2= \ell^2\;\frac{\psi'_{\alpha +}}{\psi_{\alpha+}}\frac{\psi'_{\beta -}}{\psi_{\beta-}},\qquad \alpha,\beta=1,2.
\ee
{where two bifurcate horizons are location of $r_{\alpha\beta}$ intersections.} All the geometric properties of Ba\~nados geometries may be conveniently described through $\psi_\pm$ functions. More analysis and discussion on Ba\~nados geometries may be found in \cite{Sheikh-Jabbari:2016unm, Sheikh-Jabbari:2016znt} and references therein.

For the later use we also note that \eqref{Hills} may be solved through the following ansatz:
\be\label{Psi-J}
\psi_\pm (x^\pm)=\exp{\left(\frac6c{\int^{x^\pm} J_\pm(x^\pm)}\right)}
\ee
where
\be\label{J-L}
J'_\pm+\frac{6}{c}J_\pm^2=\frac{c}{6}L_\pm.
\ee
The above equation determines $J_\pm$ for any given $L_\pm$ and has two solutions $J_{\alpha\pm}$.  {For the family of black holes, hence, $J_\pm(x^\pm)$ are real-valued, periodic functions. This among other things, recalling \eqref{J-L}, implies that $\int_0^{2\pi} L_\pm$ for black hole solutions {are} positive-definite, in accord with what is expected from the corresponding Virasoro coadjoint orbits \cite{Balog,Sheikh-Jabbari:2016unm}.} The horizon radii \eqref{horizon-radii} take a very simple form in terms of $J_{\alpha\pm}$ functions, $r_{\alpha\beta}^2=\frac{36\ell^2}{c^2}J_{\alpha +}J_{\beta -}$. In other words, functions $J_{\alpha\pm}$ determine the ``shape'' of the horizon surfaces. 

{We note that the family of locally AdS$_3$ geometries discussed in \cite{Afshar:2016wfy} are specific family of Ba\~nados geometries for which $J_\pm$ are real valued and whose value of $L_\pm$ are related to the $\gamma, \omega$ functions there through \eqref{J-L} where $J_\pm=\gamma\pm \omega$. We shall return to this point in our analysis in the next sections. This point will be further discussed and explored in an upcoming publication \cite{Upcoming}.}

Finally, we would like to mention that, if the AdS$_3$ gravity theory \eqref{AdS3-gravity-theory} has a 2d CFT dual, this CFT should be at the Brown-Henneaux central charge $c$
\be\label{central-charge}
{c = \frac{3\ell}{2G}, 
 }
\ee 
and to have a well-defined semi-classical gravity description, $c\gg 1$.
The Ba\~nados geometries correspond to (set of) states in the 2d CFT for which the expectation value of the left and right sectors of the 2d energy momentum tensor $\bT_\pm$
are given as\footnote{ {We comment that $L_\pm$ are real-valued periodic functions, \eqref{T-VEV}, while implying $J_\pm$ should be periodic, does not imply $J_\pm$ should be real. In particular, $J_0^\pm=\frac1{2\pi}\int_0^{2\pi} J^\pm(x^\pm)$ can take pure imaginary or real values \cite{Upcoming}.}}
\be\label{T-VEV}
\langle \bT_\pm \rangle =\frac{c}{6} L_\pm=\frac{6}{c} J_\pm^2+J'_\pm.
\ee

\section{The two, near horizon and asymptotic, symmetry algebras}\label{section-two-algebras}

As discussed in the introduction, depending on the choice of fall behaviour of the metric perturbations near the boundary we can have residual diffeomorphisms which respect this fall off behaviour.  Besides the near boundary (asymptotic) behaviour, for the geometries which have a horizon, one may explore the question whether we have a choice of boundary conditions which lead to nontrivial residual near horizon diffeomorphisms. If yes, we will then have (at least) two sets of residual diffeomorphisms and the corresponding charges and algebras. As we will see and discuss, however, despite being seemingly different, these two algebras and the corresponding Hilbert spaces are closely related to each. In this section we review the two algebras and their relation. In the next section we will study the relation between the corresponding Hilbert spaces.

\textbf{Note on conventions:} Through this paper we use boldface characters to denote operators and use calligraphic mathematical symbols to denote near horizon quantities. The near horizon algebra generators will hence be denoted as calligraphic-boldface. Expectation values of the operators would be denoted by the same symbol, but not boldfaced. 

\subsection{Near horizon algebra}  

As discussed one may get  nontrivial symmetries through imposing appropriate boundary conditions on the near horizon geometry of a generic non-extreme black hole. In this case one can show that the near horizon algebra  consists of  copies of the {creation-annihilation algebra} (generated by current algebra-type operators) \cite{Afshar:2016wfy}, or BMS-type algebras (with or without $u(1)$ currents) \cite{NH-algebra-1,NH-algebra-2}.
For the particular case of non-extremal BTZ black holes this algebra is two copies of current algebras \cite{Afshar:2016wfy}. It is fairly straightforward to check that the analysis of \cite{Afshar:2016wfy} readily extends to \emph{any non-extremal black hole} geometry in the Ba\~nados family. The reason is simply the fact that details of the black hole geometry do not appear near the bifurcate Killing horizon. The near horizon algebra for any black hole in the Ba\~naods family is {then} \cite{Afshar:2016uax,Afshar:2016wfy}{
\be\label{NH-algebra}
[\bcJ_n^\pm,\,\bcJ_m^\pm] =\tfrac12 \,n\, \delta_{n,-m}\,,\qquad [\bcJ_n^+,\,\bcJ_m^-]=0\;.
\ee 	

The $\bcJ_n^\pm$ generators may be viewed as creation-annihilation operators for a free 2d boson theory on $\mathbb{R}\times S^1$ which is a 2d conformal field theory (CFT$_2$) and the $\bcL_n$'s are  Fourier modes of its energy-momentum tensor:
\be\label{NH-Vir-gen}
\bcL_n^\pm\equiv\sum_{p\in\mathbb{Z}}\colon\!\bcJ_{n-p}^\pm\,\bcJ_p^\pm\colon
\ee 
where $\colon\colon$ denotes normal ordering ($\bcJ_n$ with $n>0$ may be viewed as annihilation operators). We then obtain
\be\label{NH-algebra2}
[\bcL_n^\pm,\bcL_m^\pm]=(n-m)\bcL_{n+m}^\pm+\frac{1}{12}(n^3-n)\delta_{n,-m},\qquad
 [\bcL_n^\pm,\bcJ_m^\pm]=-m \bcJ_{n+m}^\pm,
\ee 
and $[\boldsymbol{{\cal X}}^+,\boldsymbol{{\cal Y}}^-]=0$ for any $\boldsymbol{{\cal X}}, \boldsymbol{{\cal Y}}$. This is two copies of Virasoro algebra at central charge one plus a $u(1)$ current. We stress that the {``independent''}  {part of the near horizon algebra is \eqref{NH-algebra} and as we will see the Virasoro generators \eqref{NH-Vir-gen} which are constructed through the $\bcJ_n$'s, are convenient operators for the comparison to the asymptotic symmetry algebra.}

 {As some remarks, we note that (1)  the near horizon algebra \eqref{NH-algebra} is independent of the AdS$_3$ radius and the details of the black hole we started from; (2) $\bcJ_0^\pm$ are central elements in the algebra commuting with all the other generators; (3) as is seen from \eqref{NH-Vir-gen}, the central charge of the near horizon Viraroso algebra, which is one, is purely quantum in nature. In the sense that it arises from the normal ordering in the definition \eqref{NH-Vir-gen}}

\subsection{Asymptotic Virasoro algebra}

The asymptotic symmetry group of asymptotically AdS$_3$ geometries with the Brown-Henneaux boundary conditions has been analysed in \cite{Brown-Henneaux}; it is  two copies of Virasoro algebra at the Brown-Henneaux central charge \eqref{central-charge}.  {Here we discuss a specific realisation of this algebra for black holes in the family of Ba\~nados  geometries.}

\subsubsection{ {Realisation of asymptotic symmetry algebra for asymptotic AdS$_3$ black holes}}

In \cite{Afshar:2016wfy} it was noted that  {for black holes in the asymptotic AdS$_3$ geometries, the Brown-Henneaux boundary conditions instead of the AdS$_3$ boundary may also be imposed, in an appropriate coordinate system, near the horizon. Specifically, in \cite{Afshar:2016wfy} a family of locally AdS$_3$ black holes which are specified by two arbitrary real-valued periodic functions was constructed and analysed. As discussed in section \ref{section-Banados-review}, these black hole geometries are a subset of Ba\~nados geometries. These two  functions was then argued to be associated with two current algebras, which expectedly, is \eqref{NH-algebra}. As discussed in \cite{Afshar:2016uax} and will become clear in section \ref{subsection-Asymp-NH-mapping}, we find it convenient to represent this algebra in a slightly different notation and use $\bJ^\pm_n$ to denote its generators }\cite{Afshar:2016wfy}:\footnote{We would like to thank H. Afshar and D. Grumiller for clarifying explanations on this issue.}  
  \be\label{Asymptotic-algebra-J}
[\bJ_n^\pm,\,\bJ_m^\pm]=\frac{c}{12} \,n\, \delta_{n,-m},\qquad   [\bJ_n^+,\,\bJ_m^-]=0,
\ee
where  $n,m\in\mathbb{Z}$. The generators of the well-established Brown-Henneaux Virasoro algebra $\bL_n$  is then naturally obtained from $\bJ_n$ through a twisted Sugawara construction
\be\label{twisted-sugawara}
\bL_n^\pm\equiv in\bJ_n^\pm+\frac6{c}\sum_{p\in\mathbb{Z}}\bJ^\pm_{n-p}\bJ^\pm_p. 
\ee
It is straightforward to check that
\be\label{Extended-asymptotic-algebra}
[\bL_n^\pm,\,\bL_m^\pm]=(n-m)\bL_{n+m}^\pm+\frac{c}{12}\,n^3\,\delta_{n,-m},\quad
[\bL_n^\pm,\,\bJ_m^\pm]=-m\,\bJ^\pm_{n+m}+ i\frac{c}{12} \,m^2\, \delta_{n,-m},
\ee
and $[\boldsymbol{X}^+,\boldsymbol{Y}^-]=0$ for any $\boldsymbol{X}, \boldsymbol{Y}$.
\footnote{As mentioned our asymptotic algebra with generators $\bL_n^\pm, \bJ_n^\pm$ arises from the usual asymptotic Brown-Henneaux boundary conditions which allows for a holomorphic and an antiholomorphic functions. However, in the AdS$_3$ case one has the possibility of relaxing the Brown-Henneaux boundary conditions. For example, the boundary conditions allowing for boundary metric to vary up to a conformal transformation was considered in \cite{Cedric}. This relaxed boundary conditions leads to the {enhanced (asymptotic) symmetry algebra} \cite{Cedric} which is the Virasoro algebra by a $u(1)$ current where now the Virasoro and current generators are independent and not related as in \eqref{twisted-sugawara}. Note that besides the $\bL(\bJ)$ relation \eqref{twisted-sugawara}, the algebra in \cite{Cedric} is different than our algebra \eqref{Extended-asymptotic-algebra}, as the $[\bL_n,\bJ_m]$ commutator there do not involve the anomaly term proportional to $m^2$. We should also note that there are other possibilities for boundary condition leading to other algebras, e.g., see \cite{Daniel, Song-Compere-Strominger}. }

It is notable that the above is nothing but \eqref{J-L} and \eqref{T-VEV}  and that the Virasoro generators $\bL_n$ are Fourier modes of the 2d energy momentum tensor $\bT$.\footnote{{We note that  the twisted Suagawara construction is closely associated with a one-dimensional linear dilaton background string theory \cite{Afshar:2016wfy}. The string worldsheet field comes from the conformal factor of the 2d $x^\pm$ part of the 3d metric. Intriguingly, the  ``string tension'' $\alpha'$ of this theory is proportional to central charge $c$ (in the semiclassical large $c$ limit) and the linear dilaton background arises from the cosmological constant of the 3d geometry.\label{footnote-linear-dilaton}}} In other words, the functions $\psi_\pm$ are related to the expectation value of the Wilson line operator (of the $sl(2,\mathbb{R})\times sl(2,\mathbb{R})$ Chern-Simons formulation of AdS$_3$ gravity) made from the holomorphic current fields $J_\pm(x^\pm)$ \cite{Afshar:2016wfy}. Note also that as \eqref{horizon-radii} shows, the equation defining the bifurcate horizons are directly related to the product of $J^\pm$ currents.\footnote{{We note that one may still enhance this algebra 
by the addition of two more generators $\mathbf{Q}^\pm$ which are ``conjugate'' to  $\bJ_0^\pm$, i.e. $[\bJ_n^\pm,\mathbf{Q}^\pm]=\frac{ic}{12} \delta_{n,0},\ [\bL_n^\pm, \mathbf{Q}^\pm]= i\bJ_n^\pm$ \cite{ASS-work-in-progress}. The $\mathbf{Q}^\pm$ were also noted in the eq.(7.9) \cite{Cedric}, where the $Q$ generators there is related to $\mathbf{Q}^\pm$ as $Q=\mathbf{Q}^+ +\mathbf{Q}^-$.\label{footnote-3}}}

 $\bJ_0^\pm$ is a central element of the algebra and commutes with all $\bJ_n^\pm$ and $\bL_n^\pm$'s. This observation makes a direct connection with the surface charge analysis of \cite{Compere:2015knw, Hajian:2015xlp}, where it was argued that the charges associated with the two exact symmetries (the two periodic Killing vectors) should commute with the Virasoro generators. Explicitly, one may identify the two exact symmetry charges (which in \cite{Compere:2015knw, Sheikh-Jabbari:2016unm} was denoted by $J_\pm$) as a function of the center element $\bJ_0^\pm$, as we will discuss below $J_\pm\propto (\bJ_0^\pm)^2$. 

\subsubsection{``Asymptotic'' and ``near horizon'' symmetry algebras are defined everywhere }\label{section-everywhere} The usual way of identifying residual symmetries and the associated charge algebras is through imposing certain boundary conditions in the asymptotic region of spacetime. However, there are cases where one can provide alternative ways to identify the symmetries. Technically, and for the example of Ba\~nados geometries, one can view this set of solutions as a phase space with a given symplectic two-form. Then, the conserved charge would be related to transformations on the phase space which do not change the {symplectic} two-form. In this case, the ``asymptotic'' symmetries become ``symplectic'' symmetries. For the case of Ba\~nados geometries this has been demonstrated in \cite{Compere:2015knw}. For other cases of symplectic symmetries see \cite{Hajian:2015xlp, symplectic,  Mitchell}.

Promoting the asymptotic symmetries to symplectic symmetries means that the conditions defining these charges (like ``the fall off behaviour'') and the surface integrals, integrals over  co-dimension two compact space-like surfaces, can be chosen or specified not only in the asymptotic region but at any place in spacetime. However, one should make sure that there are no topological obstructions, such as singularities, or physical obstructions, like closed-time-like curves (CTC's), in the spacetime. Dealing with symplectic symmetries means that the algebra \eqref{Asymptotic-algebra-J} and \eqref{Extended-asymptotic-algebra} can be defined everywhere in the geometry.\footnote{For the case of 3d gravity, where we have a Chern-Simons description, the fact that charges may be defined at any radius $r$, and are $r$-independent can be seen very explicitly from the fact that $r$-dependence, in the Chern-Simons perturbations can be removed by a gauge choice, e.g., see \cite{NH-algebra-2, Afshar:2016wfy, Daniel, Banados-94, Banados:1998ta, Afshar:2014rwa}.} 
{Moreover, one can show \cite{ASS-work-in-progress} that the algebra \eqref{NH-algebra} or equivalently \eqref{Asymptotic-algebra-J} is the symplectic symmetry of the family of black  hole geometries discussed in \cite{Afshar:2016wfy}.}
Despite these facts and to distinguish this algebra from the one appearing on the near horizon, we keep calling them as ``asymptotic'' symmetries.

\subsection{Asymptotic algebra from  near horizon algebra}\label{subsection-Asymp-NH-mapping} 

 {As discussed the near horizon algebra \eqref{NH-algebra} or \eqref{Asymptotic-algebra-J}, and the Brown-Henneaux Virasoro algebra \eqref{Extended-asymptotic-algebra} are both symplectic symmetries of the phase space of locally AdS$_3$ black hole solutions  and may hence be defined everywhere in the corresponding coordinate systems \cite{Afshar:2016wfy, Compere:2015knw, ASS-work-in-progress}. Moreover, from $\bcJ _n^\pm$ one construct $\bcL_n^\pm$ \eqref{NH-Vir-gen} which generate a Virasoro algebra at central charge one.}

 {In \cite{Banados-map} there is a simple proposal how to relate two  Virasoro algebras at central charge $c$ and at central charge one. Motivated by the proposal in \cite{Banados-map}, we related the associated $\bL_0^\pm, \bcL_0^\pm$ as
\be\label{L0-map}
c \bL_0^\pm=\bcL_0^\pm-\frac1{24}.
\ee 
The extra $-1/24$ term in the RHS has appeared because of difference between the two conventions used in the ``near horizon'' \eqref{NH-Vir-gen} and ``asymptotic'' \eqref{Extended-asymptotic-algebra} Virasoro algebras. The relation \eqref{L0-map}, while could be (partially) justified with the arguments in \cite{Banados-map}, is a part of our horizon fluffs proposal. }

 {We next require that the relation \eqref{L0-map} is consistent with the commutation relations among other generators. This requirement then yields \cite{Afshar:2016uax}
\footnote{Recalling that the two algebras \eqref{NH-algebra} and \eqref{Asymptotic-algebra-J} are the same, up to possibly normalisation of charges which is not fixed through the usual surface charge computation methods, we get a relation of the form 
$$
\bJ_n^\pm =\sqrt{\frac{c}{6x}}\bcJ_{x n}^\pm,\qquad n\neq 0,
$$
where $x$ can be any arbitrary integer. The $x=c$ choice is fixed upon imposing \eqref{L0-map}.} 
\be\label{NH-infty-map}
\bJ_n^\pm =\frac{1}{\sqrt{6}}\bcJ_{c n}^\pm,\qquad n\neq 0.
\ee
In the above we have assumed central charge $c$ is integer-valued. This assumption is a well justified one, noting the discussions of \cite{Witten-AdS3} and that the modular invariance of a presumed unitary 2d CFT dual to the pure AdS$_3$ Einstein gravity  implies the central charge $c$ to be a multiple of 24.\footnote{Analysis of \cite{Witten-AdS3} casts some doubts on having a well-defined pure AdS$_3$ \emph{quantum} gravity (which is dual to a unitary-modular invariant 2d CFT).  Nonetheless,  we are working in the large $c$ regime where there is a semiclassical gravity description and integer $c$ assumption is a justified one in this limit.}
Eq.\eqref{NH-infty-map} does not relate $\bcJ_0^\pm$ and $\bJ_0^\pm$. The relation between these two is induced from \eqref{L0-map}:
\be\label{J0-map}
\bL_0^\pm=\frac{1}{c}(\bcL_0^\pm-\frac1{24})  \quad \Rightarrow \quad 6(\bJ_0^\pm)^2=(\bcJ_0^\pm)^2+\sum_{n\neq 0}\left(:\bcJ_{-n}^\pm \bcJ_{n}^\pm:-\sum_{n\neq 0}\bcJ_{-nc}^\pm \bcJ_{nc}^\pm\right)-\frac1{24},
\ee
where we used \eqref{twisted-sugawara} and \eqref{NH-Vir-gen}. Given the above one can find other $\bL_n^\pm$ in terms of the near horizon algebra generators $\bcJ^\pm_n$.}

\section{Hilbert space and representations of the two algebras}\label{section-Hilbert-spaces}

 {As already mentioned our symmetry algebras \eqref{NH-algebra} and \eqref{Extended-asymptotic-algebra} are respectively (symplectic) symmetries of the family of metrics discussed in \cite{Afshar:2016wfy} and \eqref{generic-Banados-geometry}. These family of solutions may hence be viewed as representations or coadjoint orbits of the associated algebras. In a more quantum language, 
one may view them as (a part of) Hilbert space of the associated physical theory which is invariant under the corresponding algebra. Besides this geometric picture, we can directly construct the Hilbert space of the two algebras and study their relation. That is what we carry out in this section.
}

\subsection{Asymptotic black hole Hilbert space $\cH_{\cB\cH}$ and black holes in Ba\~nados geometries}\label{subsubsection-4.1}

The asymptotic algebra involves a Virasoro algebra at Brown-Henneaux central charge and hence Virasoro coadjoint orbits may be used to label Ba\~nados geometries. This has been studied in some detail in \cite{Sheikh-Jabbari:2016unm, Compere:2015knw} and we do not review that here. What we would like to note here is that realisation of Virasoro generators in terms of $u(1)$ currents $\bJ_n^\pm$, i.e. the twisted-Sugawara map \eqref{twisted-sugawara},  {implies that only a subset of Virasoro coadjoint orbits associated with AdS$_3$ black holes, namely the hyperbolic orbits (see \cite{Balog,Sheikh-Jabbari:2016unm}) appear as coadjoint orbits of our $\bL_n^\pm,\bJ_n^\pm$ algebra. }

 {To see the above explicitly, and for our later use, we construct these Virasoro orbits using the current algebra.} To this end, we note that $\bJ_n, n<0$ and $\bJ_n, n>0$ are respectively creation annihilation conjugates of each other. Therefore, it is natural to define the ``asymptotic black hole vacua'' $|0; J_0^\pm\rangle_{_{{\cB\cH}}}$ as
\be\label{Asymp-vacuum-def}
\bJ_n^\pm |0; J_0^\pm\rangle_{_{{\cB\cH}}}=0,\qquad  \forall n>0, \qquad \bJ_0^\pm |0; J_0^\pm\rangle_{_{{\cB\cH}}}=J_0^\pm|0; J_0^\pm\rangle_{_{{\cB\cH}}},
\ee
and we may  choose the normalisation such that
\be\label{J-state-normalization}
{}_{_{\cB\cH}}\langle 0; {J'}_0^\pm|0; J_0^\pm\rangle_{_{{\cB\cH}}}=
\delta_{{J'}^\pm_0,J^\pm_0}.
\ee
 {Analysis of \cite{Afshar:2016wfy} show that $J_0^\pm$ for set of AdS$_3$ black hole geometries discussed there  should be real valued, which without loss of generality may be chosen to take only positive values, i.e. $J_0^\pm\in \mathbb{R}^+$. This latter implies that $\bJ_0^\pm$ is self-adjoint (hermitian) on the Hilbert space associated with this set of geometries. However, if we ignore the analysis of \cite{Afshar:2016wfy} and only focus on the algebra \eqref{Asymptotic-algebra-J} or \eqref{twisted-sugawara}, there is no hermiticity condition induced on $\bJ_0^\pm$ and $J_0^\pm$ can take real or imaginary values. We will return to this point, shortly in section \ref{Vir-orbits-HA}. We define the asymptotic Hilbert space ${\cH}_{{\cB\cH}}$ as the states constructed upon vacuum states $|0; J_0^\pm\rangle_{_{{\cB\cH}}}$ with real-positive $J^\pm_0$.}

{The rest of states in the asymptotic Hilbert space ${\cH}_{{\cB\cH}}$ are then constructed by the action of $\bJ_{-n}^\pm, n>0$ on the vacuum. That is,   ${\cH}_{{\cB\cH}}$ consists of states of the form}
\be\label{generic-NH-state}
\begin{split}
 |{{\lbJ}}(\{n_i^\pm\}); J_0^\pm\rangle_{_{{\cB\cH}}} &=  {{\lbJ}}(\{n_i^\pm\}) |0; J_0^\pm\rangle_{_{\cB\cH}}, \qquad \forall |{{\lbJ}}(\{n_i^\pm\}); J_0^\pm\rangle_{_{\cB\cH}} \in {\cH}_{\cB\cH},\\
 {{\lbJ}}(\{n_i^\pm\})&={\cal N}_{{\bJ}}\prod_{\{n_i^\pm>0\}}\!\!\big( \bJ_{-n_i^+}^+\cdot \bJ_{-n_i^-}^-\big) ,
\end{split}
\ee
where  the normalisation factor ${\cal N}_{{\bJ}}$ is fixed such that 
\be
{}_{_{\cB\cH}}\langle {{\lbJ}}(\{{n'}_i^\pm\}); {J'}_0^\pm|{{\lbJ}}(\{n_i^\pm\}); {J}_0^\pm\rangle_{_{{\cB\cH}}}=\delta_{\{{n}'_i\},\{{n}_i\}}\delta_{{J'}^\pm_0,J^\pm_0},
\ee
where $\delta_{\{{n}'_i\},\{{n}_i\}}$ is the product of all $\delta_{n'_i,n_i}$'s.

\subsubsection{ {${\cH}_{\cB\cH}$ vs. Hilbert space of unitary Virasoro (coadjoint) orbits $\cH_{Vir}$} }\label{Vir-orbits-HA}
Given a generic state, the associated Virasoro orbit may be constructed by the action of all possible products  of $\bL_{-n}, n>0$ on this state. Since we know the commutation of $\bL_n,\ \bJ_m$ we can readily distinguish Virasoro orbits in ${\cH}_{\cB\cH}$. In particular, since
\be
[\bJ_0^\pm, \bL_n^\pm]=0,
\ee
 all states in the same orbit  have the same $J^\pm_0$. In other words, the orbits may be labelled by $J^\pm_0$, or $J^\pm_0$ is an \emph{orbit invariant quantity}.\footnote{We comment that  $J^\pm(x^\pm)= J^\pm_0+ \sum_{n\neq 0} J^\pm e^{inx^\pm}$ and \eqref{Psi-J} imply  $\psi^\pm= \exp(\frac{6}{c}J_0^\pm x^\pm) P^\pm(x^\pm)$, where $P^\pm$ are periodic functions. In other words, $\frac{6}{c}J^\pm_0$ are equal to the Floquet index, denoted by ${\cal T}_\pm$  in \cite{Sheikh-Jabbari:2016unm} (\emph{cf.} foonote \ref{Footnote-Floquet}). } That is, there is an orbit associated with each $|0; J_0^\pm\rangle_{_{{\cB\cH}}}$ state, or, $|0; J_0^\pm\rangle_{_{{\cB\cH}}}$ is the representative of an orbit \cite{Balog}. Recalling \eqref{twisted-sugawara}, one observes that 
\be
\bL_0^\pm |0; J_0^\pm\rangle_{_{{\cB\cH}}}\equiv L_0^\pm |0; J_0^\pm\rangle_{_{{\cB\cH}}} = \frac{6}{c}(J^\pm_0)^2 |0; J_0^\pm\rangle_{_{{\cB\cH}}}.
\ee
The condition for unitarity of the representation implies hermiticity of $\bL_0^\pm$ and  $L_0^\pm+ \frac{c}{24}\geq 0$. In other words, if we only focus on the Virasoro part of the algebra, one may relax hermiticity of $\bJ^\pm_0$ and require hermiticity of $(\bJ_0^\pm)^2$ and $(J^\pm_0)^2+(\frac{c}{12})^2\geq 0$.

Recalling discussions in sections 3,4 of \cite{Sheikh-Jabbari:2016unm}, that the Ba\~nados geometries are in one-to-one relation with the Virasoro coadjoint orbits and that the orbits discussed here are in one-to-one relation with the coadjoint orbits, the above implies that  $\cH_{\cB\cH}$ which only includes  real-valued $J_0^\pm$ only captures  hyperbolic orbits associated with BTZ black holes. If we allow for negative values of $(J_0^\pm)^2$, those with $-(\frac{c}{12})^2<(J^\pm_0)^2<0$ capture elliptic orbits, associated with conic singularities on global AdS$_3$. The ``global AdS$_3$ vacuum'' has $(J^\pm_0)^2 =-(\frac{c}{12})^2$ \cite{Sheikh-Jabbari:2016unm}.

 {Therefore, the set of all unitary of representations of Virasoro algebra and their coadjoint orbits form a bigger Hilbert space than $\cH_{\cB\cH}$. This Hilbert space will conveniently be denoted by $\cH_{Vir}$ and contains orbits associated with global AdS$_3$ $(|0; J_0^\pm=ic/12\rangle)$, elliptic orbits associated with $(|0; J_0^\pm=ic\nu^\pm/12\rangle), 0<\nu^\pm<1$, and hyperbolic orbits associated with $(|0; J_0^\pm\rangle_{_{{\cB\cH}}}), J^\pm_0\in\mathbb{R}^+$ (BTZ black hole orbits). Therefore, it is readily seen that  $\cH_{\cB\cH}\subset \cH_{Vir}$.}

Before closing this part some further comments are in order:\vskip -5mm
\begin{itemize}\vskip -5mm
\item[\textit{i.}] All states in the same Virasoro orbit will have the same $J_0^\pm$ while they will have different $\bL_0^\pm$ eigenvalues $L_0^\pm$. Explicitly, $L_0^\pm$ is \emph{not} an orbit invariant quantity.
 \item[\textit{ii.}] States in the orbit of $|0; J_0^\pm\rangle_{_{{\cB\cH}}}$ all have higher $\bL_0^\pm$ eigenvalue than $|0; J_0^\pm\rangle_{_{{\cB\cH}}}$ itself. That is, the representative  (which may be called ``highest weight state'') has the lowest value of $L_0^\pm$ in the orbit.
 \item[\textit{iii.}] 
 { There could be more than one orbit associated with a given $J^\pm_0$, as the orbits can have another \emph{discrete} label. This other label, being discrete, is not captured by the surface integrals accounting for charges associated with continuous transformations, and hence not included in our current discussions. These other orbits, however, are not among the representations which are unitary on a single cover of AdS$_3$, while they may be unitarisable on multiple covers of AdS$_3$. (See \cite{Sheikh-Jabbari:2016unm} for some preliminary discussion on the latter.) These other cases may be studied in a separate publication.}
 
 \end{itemize}

\subsubsection{Generic black hole is a coherent state in $\cH_{\cB\cH}$}\label{subsection-coherent}

As mentioned in the previous subsections, the asymptotic Hilbert space $\cH_{\cB\cH}$ contains states with $J_0^\pm\in \mathbb{R}^+$ associated with geometries  in the orbit of BTZ black holes (related to hyperbolic orbits). Geometries in these orbits are specified by generic functions $L_\pm(x^\pm)$ which can be mapped to constant-positive $L^\pm_0$, explicitly, there exists 
$h_\pm(x^\pm)$ functions such that \cite{Sheikh-Jabbari:2016unm}
$$
L_\pm(x^\pm)={h'_\pm}^{2} L^\pm_0 + S[h_\pm;x],\qquad h_\pm(x^\pm+2\pi)=h_\pm(x^\pm)+2\pi,
$$
where $S[h;x]$ is the Schwarz derivative.\footnote{Recalling \eqref{Psi-J}, one may see that under a conformal transformation $J(x)$ behaves as: $J(x)\to J(h(x)) h'-\frac{c}{12}\frac{h''}{h'}$, compatible with $\bL, \bJ$ commutator in \eqref{Extended-asymptotic-algebra}.} Alternatively, as reviewed in section \ref{section-Banados-review} (see also \cite{Sheikh-Jabbari:2016unm, Sheikh-Jabbari:2016znt}) one may use $\psi_\pm$, or equivalently $J_\pm(x^\pm)$, to specify the corresponding geometry. Explicitly, let us consider a geometry corresponding to the (coadjoint) orbit associated with the BTZ geometry with given $J^\pm_0$. This geometry is then associated to  a
state in $\cH_\cB\cH$  which we conveniently denote by  $|\{J_{n_i}^\pm\}; J_0^\pm\rangle_{_{{\cB\cH}}}$. This state is defined through equation
\be\label{Banados-asymptotic}
{}_{_{{\cB\cH}}}\langle \{J_{n_i}^\pm\}; J_0^\pm|\bJ_{n_k}^\pm |\{J_{n_i}^\pm\}; J_0^\pm\rangle_{_{{\cB\cH}}}=J_{n_k}^\pm,\qquad |\{J_{n_i}^\pm\}; J_0^\pm\rangle_{_{{\cB\cH}}}\in\cH_{\cB\cH},
\ee
where the $\{J^\pm_{n_i}\}$ appearing in $|\{J_{n_i}^\pm\}; J_0^\pm\rangle_{_{{\cB\cH}}}$ is the collection of all $J^\pm_{n_k}$ appearing in the right-hand-side.
The BTZ black hole discussed in \cite{Afshar:2016uax} corresponds to $J_n^\pm=0, (n\neq 0)$ in the above family. Since in the BTZ orbit we are dealing with geometries with real $J_\pm(x^\pm)$ functions, we have
$J_{-n}=J_n^*$.

Recalling the fact that each $\bJ_n$ behaves like creation ($n<0$) or annihilation ($n>0$) operator, solutions to \eqref{Banados-asymptotic} are coherent states. Next, we also note that $\bJ_n, \bJ_m$ commute for $n\neq -m$ and hence a  general solution to set of equations \eqref{Banados-asymptotic} for all $n$ is simply product of these coherent states. That is,
{\be\label{coherent-state-BTZ-family}
 |\{J_{n_i}^\pm\}; J_0^\pm\rangle_{_{{\cB\cH}}}= \!\!\prod_{\{n_i^\pm>0\}} \exp{(\frac{12J^+_{n_i}}{c\ n_i} \bJ_{-n_i}^+ -\frac{{12J^+_{n_i}}^*}{c\ n_i} \bJ_{n_i}^ +)}\cdot  \exp{(\frac{12J^-_{n_i}}{c\ n_i} \bJ_{-n_i}^- -\frac{12{J^-_{n_i}}^*}{c\ n_i} \bJ_{n_i}^-)} |0; J_0^\pm\rangle_{_{\cB\cH}}. 
 \ee}
One may readily check that the above coherent state is normalised ${}_{_{{\cB\cH}}}\langle\{J_{n_i}^\pm\}; J_0^\pm|\{J_{n_i}^\pm\}; J_0^\pm\rangle_{_{{\cB\cH}}}=1$.

As we see the black hole state is not a coherent state in $\bJ^\pm_0$ sector. This is due to the fact that $\bJ_0^\pm$, unlike $\bJ_n^\pm, \bJ_{-n}^\pm$, have no conjugate in the algebra \eqref{Asymptotic-algebra-J}. However, as pointed out in footnote \ref{footnote-3}  one may enhance \eqref{Asymptotic-algebra-J} by the addition of $\bQ^\pm$ operators which are conjugate to $\bJ_0^\pm$.  We will briefly comment on this point and its possible implications for our black hole microstate counting in the discussion section.

Finally, we remark that using \eqref{twisted-sugawara} one can read the values of $L_n^\pm$ for the BTZ family in terms of their $J^\pm_n$ eigenvalues, namely we have
\be\label{Ln-Banados-asymptotic}
{}_{_{{\cB\cH}}}\langle \{J_{n_i}^\pm\}; J_0^\pm|\bL_n^\pm |\{J_{n_i}^\pm\}; J_0^\pm\rangle_{_{{\cB\cH}}}=L_n^\pm ,
\ee
where $L^\pm_n=inJ^\pm_n+\frac{6}{c}\sum J^\pm_{n-p}J^\pm_p$. As discussed values of $L_n^\pm$ and in particular $L_0^\pm$ are not orbit invariant quantities and the appropriate definition of mass or angular momentum, as inspired by the gravity analysis, is through orbit invariant quantities $J^\pm_0$. This latter leads to the conserved quantities associated with the  two $U(1)$ global Killing vectors of Ba\~nados geometries and are those appearing in the expression for the first law of thermodynamics \cite{Sheikh-Jabbari:2016unm,Hajian:2015xlp}. For the special case of BTZ black hole, where all $J_n^\pm, n\neq 0$ vanish, then (\emph{cf.} \eqref{T-VEV}, \eqref{twisted-sugawara})
\be\label{BTZ}
\text{For BTZ geometry:} \qquad  \frac12(\ell M\pm J)\equiv  \Delta^\pm=\frac{6}{c}(J_0^\pm)^2.
\ee 
where $M, J$ are ADM mass and angular momentum of the BTZ black hole.

\subsection{ {Hilbert space of near horizon soft hairs $\HNH$}}

As discussed in \cite{Sheikh-Jabbari:2016unm} the causal structure of all Ba\~nados geometries in the same Virasoro (coadjoint) orbit are the same. Therefore, despite the fact that in the BTZ orbits we are dealing with geometries with non-constant value of $L_\pm(x^\pm)$ (i.e. they have non-zero $L^\pm_n$ values), they have the same causal structure as the corresponding BTZ. One  may hence study their near horizon algebra as was done for BTZ black hole itself. Analysis of \cite{Afshar:2016uax, Afshar:2016wfy} readily goes through and one obtains the same asymptotic and near horizon algebras \eqref{Extended-asymptotic-algebra} and \eqref{NH-algebra} for all members in the BTZ orbits.

To construct the Hilbert space of the near horizon algebra $\HNH$, we follow essentially the same steps as the previous subsection.   {We start with defining the vacuum $|0\rangle$:}
\be\label{vacuum-state-def}
 \begin{split}
 \bcJ_n^\pm|0\rangle= 0, \qquad\forall\ n\geq 0. 
\end{split}\ee
We note the difference between the above and \eqref{Asymp-vacuum-def} and that here we set $\bcJ^\pm_0$ to have zero eigenvalue on the vacuum. In principle we could have chosen the vacuum to have a non-zero $\cJ^\pm_0$ value.
 $\cJ^\pm_0=0$ is a convenient choice as $\bcL_0^\pm |0\rangle = 0$.  {More importantly, it is a convenient choice because the energy in the near horizon observer frame  is proportional to $\cJ_0^++\cJ_0^-$ \cite{Afshar:2016wfy}. With this choice, all the state in $\HNH$ will have zero energy from a near horizon observer viewpoint, they are \emph{near horizon soft hairs} \cite{HPS, Afshar:2016uax}. In other words, $\HNH$ is nothing but the Hilbert space of near horizon soft hairs.}

 
A generic descendant of the vacuum,  $|\Psi(\{n_i^\pm\})\rangle$, may then be constructed using creation operators  $\cJ_{-n_i^\pm}^\pm$ with sets of positive integers $\{n^\pm_i>0\}$, i.e.
\be\label{generic-NH-state}
 \new{|{\Psi}(\{n_i^\pm\})\rangle = \!\!\prod_{\{n_i^\pm>0\}}\!\!\big( \bcJ_{-n_i^+}^+\cdot \bcJ_{-n_i^-}^-\big) |0\rangle\,,}\qquad \forall |\Psi (\{n_i^\pm\})\rangle \in \HNH.
\ee
Given \eqref{vacuum-state-def}, we deduce that
\be\label{NH-energy}\begin{split}
\bcL_0^\pm|\Psi(\{n_i^\pm\})\rangle = \big(\sum_i  n_i^\pm \big)|\Psi(\{n_i^\pm\})
\rangle. 
\end{split}\ee
One may compute the eigenvalue of $\cL_n$'s ($n\neq 0$) and construct the near horizon (Virasoro) orbits. However, since it is essentially the same as the asymptotic algebra case and will not be relevant for our microstate counting, we will not present it here.
 {
\subsection{Relation between $\cH_{\cB\cH}$, $\cH_{Vir}$ and $\HNH$}
So far, we have introduced and constructed  three Hilbert spaces: 
\bi
\item $\cH_{\cB\cH}$ which is the Hilbert space of generic AdS$_3$ black holes, this includes all non-extremal BTZ black holes and their Virasoro excitations (descendants). 
\item $\cH_{Vir}$ which is the Hilbert space associated with all unitary constant representative Virasoro coadjoint orbits, including BTZ black holes and conic singularities and their Virasoro descendants. Every state in $\cH_{Vir}$ is then specified by the value of $J^\pm_0$ and a set of integers (accounting for Virasoro excitations). Note in particular that $\cH_{\cB\cH}\subset\cH_{Vir}$. 
\item $\HNH$ which denotes Hilbert space of near horizon soft hairs and its states are specified by a set of integers.
\ei}
 {
With the maps introduced in section \ref{subsection-Asymp-NH-mapping}, we related a subset of the near horizon generators $\bcJ_n^\pm$ to the asymptotic ones $\bJ^\pm_n$. Here we would like to discuss the relation  between the Hilbert spaces induced from this mapping of generators. In particular, \eqref{NH-infty-map} and \eqref{L0-map} or \eqref{J0-map} may be viewed as an equation which could be written on the $\HNH$. Using them we learn that the $|0; J_0^\pm\rangle_{_{{\cB\cH}}}$ vacuum state is mapped onto several different states created by $\bcJ^\pm_r$ or product thereof, with $r<c$ and with $6(J_0^\pm)^2=\sum_{r_i<c} r_i^\pm-1/24$. Therefore, we learn that $\HNH$ contains all the states in $\cH_{\cB\cH}$ whose $6(J^\pm_0)^2$ is of the form given here and $\HNH$ does not include states with other form of $J_0$'s.  These are the state which in \cite{Afshar:2016uax} were identified as BTZ black hole microstates. See next section for more discussions.}

 {On the other hand, from discussions of section \ref{section-everywhere} and the standard AdS$_3$/CFT$_2$ and Brown-Henneaux analysis, we expect $\HNH\subset\cH_{Vir}$. This expectation will be discussed and established in \cite{Upcoming}. However, here we would like to give the general picture. Given \eqref{NH-infty-map}, one may distinguish three classes of creation operators among $\bcJ^\pm$'s: $\bcJ^\pm_{-nc}$ with $n>0$, $\bcJ^\pm_{-r}$ with $r=1,2,\cdots, c-1$,  and $\bcJ^\pm_{-(nc+r)}, n>0,  0<r<c $. The first class, as discussed, are those which create states in the $\cH_{\cB\cH}$. The second and third classes, which correspond to states with imaginary $J_0^\pm$ in $\cH_{Vir}$ are then associated with certain conic defects and their Virasoro descendants. In particular, the second class are associated with  conic defects with $L_0=-1+r/c$ where $r=0$ case creates global AdS$_3$ vacuum and $r=c$ the massless BTZ state. These states are then creating the spectral flow between the two vacua. This picture dovetails with the well-known spectral flow of the {(supersymmetric)} 2d CFT's which are dual to AdS$_3$ gravity theory \cite{Spectral-flow-1,Spectral-flow-2}. 
}

\section{Microstates of a generic AdS$_3$ black hole}\label{section-BH-microstates}

The most natural proposal for black hole microstates, which was put forward in \cite{Afshar:2016uax}, is that microstates of a black hole specified by a given set of $L_n^\pm$ (or equivalently $J^\pm_N$) charges are states in $\HNH$ which describe a geometry with the same asymptotic charges $L_n^\pm$. To this end, we note that using the map \eqref{NH-infty-map} and also definitions of near horizon and asymptotic vacua, generators of the asymptotic algebra and in particular $\bJ^\pm_n$ can now be viewed as operators defined on the whole $\HNH$ (and not just on $\cH_{\cB\cH}$). Then, our proposed definition of black hole microstates associated with the Ba\~nados geometry given by function $J_\pm(x^\pm)$ with Fourier modes $J_n^\pm$ is simply equation \eqref{Banados-asymptotic}, but now this equation is to be solved over $\HNH$. Explicitly, our microstates $|\cB(\{J_n^\pm\}; J^\pm_0)\rangle\in \HNH$ are solutions to 
\be\label{Microstates-definition}
\langle\cB'(\{J_n^\pm\}; J^\pm_0)| \bJ^\pm_n|\cB(\{J_n^\pm\}; J^\pm_0)\rangle=\frac{1}{\sqrt{6}}\ \langle\cB'(\{J_n^\pm\}; J^\pm_0)| \bcJ^\pm_{nc}|\cB(\{J_n^\pm\}; J^\pm_0)\rangle= J_n^\pm \delta_{\cB,\cB'},\quad n\neq 0.
\ee

Recalling the relations between asymptotic and near horizon generators \eqref{NH-infty-map} and that $\bcJ_n, \bJ_m$ with $n<c$ commute, solution to the above is of the form, 
\be\label{Microstate-solution}
|\cB(\{J_n^\pm\}; J^\pm_0)\rangle= {\cal N}_{\{n_i^\pm\}} |\cF(\{n_i^\pm\});J^\pm_0\rangle \otimes |\{J_{n_i}^\pm\}; J_0^\pm\rangle_{_{{\cB\cH}}} 
\ee{
where
\be|{\cal N}_{\{n_i^\pm\}}|^2 =\prod_{\{0<n_i^\pm<c\}}\frac{2}{n_i^+}\cdot\frac{2}{n_i^-},\nonumber\ee}
and
\be\label{BTZ-microstates}
|\cF(\{n_i^\pm\}); J^\pm_0\rangle = \prod_{\{0<n_i^\pm<c\}} \!\!\!\!\big(\bcJ_{-n_i^+}^+ \cdot \bcJ_{-n_i^-}^-\big) |0\rangle \,,\quad \text{such that}\quad \sum n_i^\pm= 6(J^\pm_0)^2+\frac{1}{24},
\ee
or linear combinations thereof.  In fact $|\cF(\{n_i^\pm\})\rangle$ is exactly the  microstates (horizon fluffs) of the BTZ black hole \cite{Afshar:2016uax}. Hereafter, we will denote the set of states in \eqref{BTZ-microstates} by $\cH_\cF$. It is readily seen that $\cH_{\cF}$ forms a vector space and that  $\cH_{\cB\cH}$ and $\cH_\cF$ form two subspaces of $\HNH$ and overlap only on the vacuum state, as depicted in Figure \ref{Figure}.
\begin{figure}[t]
\begin{center}
\includegraphics[scale=.45]{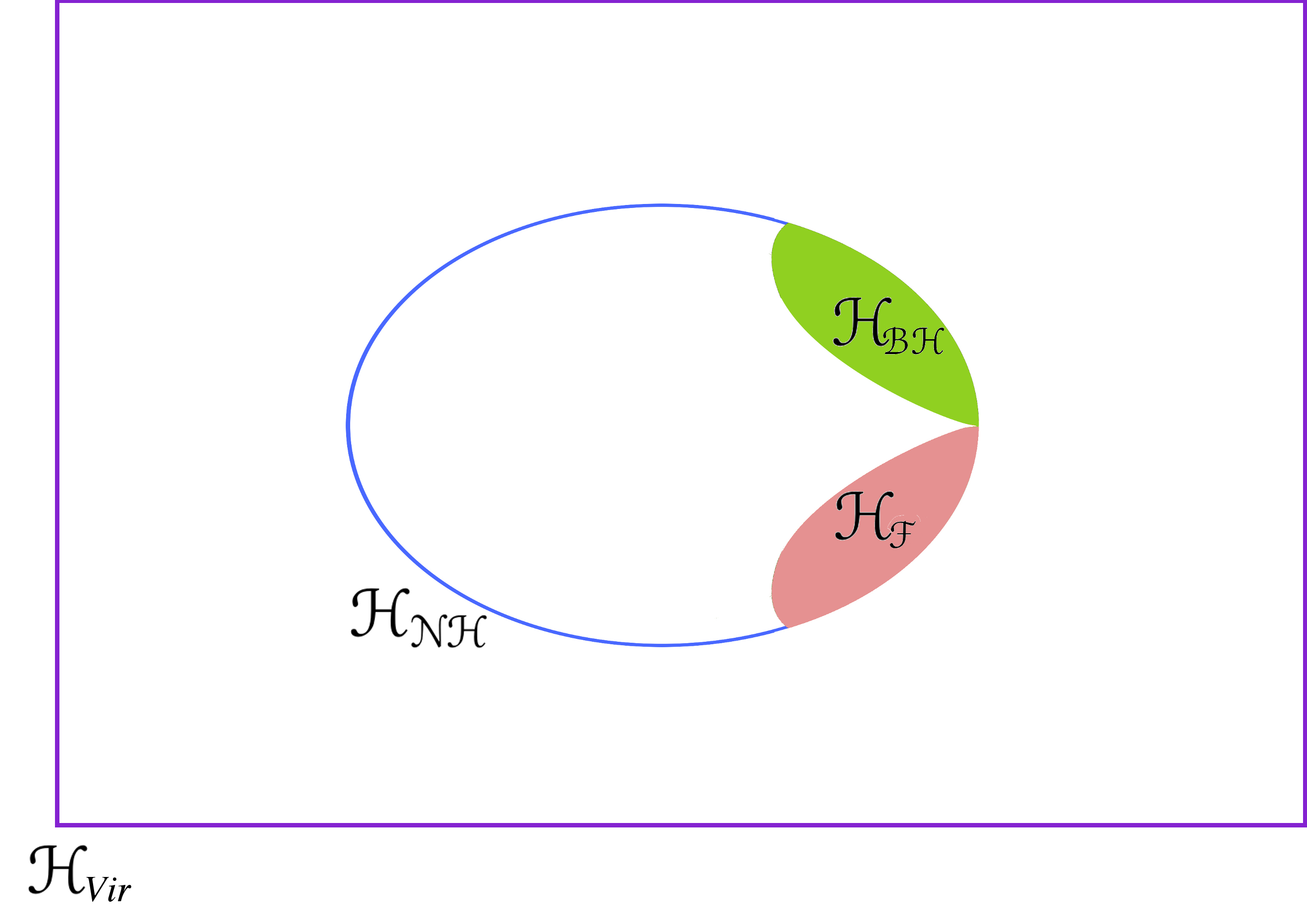}
\caption{A schematic diagram of the near horizon Hilbert spaces ${\mathcal H}_{\cN\cH}$, Asymptotic black hole Hilbert space is ${\mathcal H}_{\cB\cH}$  (green region) and the Hilbert space of horizon fluffs (black hole microstates) is $\cH_\cF$ (red region). Intersection of $\cH_{\cB\cH}$ and $\cH_\cF$ is the vacuum state $|0\rangle$ or $|0;0\rangle_{_{\cB\cH}}
$}\label{Figure}
\end{center}
\end{figure}

One may associate an orbit invariant mass and angular momentum to the BTZ orbits. This is possible recalling \eqref{T-VEV} and \eqref{twisted-sugawara}:
\be\label{orbit-invariant-mass}
\Delta^\pm=\frac12(\ell M\pm J)=\frac{c}{6}\langle\cF(\{n_i^\pm\}); J^\pm_0|\bL_0^\pm|\cF(\{n_i^\pm\}); J^\pm_0\rangle = \frac{6}{c} (J^\pm_0)^2\simeq \frac1c \sum n_i,
\ee
 {where in the last equality we assumed $\sum n_i\gg 1/24$ and dropped the 1/24 part.}
Note that here $n_i<c$, that in the above the ``BTZ part'' of the black hole microstates, $|\cF(\{n_i^\pm\}); J^\pm_0\rangle$ (and not $|\cB(\{J_n^\pm\}; J^\pm_0)\rangle$) has been used, and that  $\langle\cF(\{n_i^\pm\}); J^\pm_0|\bJ^\pm_{n\neq 0}|\cF(\{n_i^\pm\}); J^\pm_0\rangle=0$.

\subsection{Black hole microstate counting} 

Given the black hole microstates \eqref{Microstate-solution} corresponding to a generic geometry in the BTZ-orbit, we can perform their counting for a given asymptotic orbit invariant mass and angular momentum \eqref{orbit-invariant-mass}.
The idea, as spelled out in \cite{Afshar:2016uax}, is to use Hardy--Ramanujan formula, see e.g.~\cite{Carlip:1998qw} and Refs.~therein: number of partitions of a given positive integer $N$ into non-negative integers, $p(N)$, for large $N$ is given by
\be\label{Hardy-Ramanujan}
p(N)\big|_{N\gg 1} \simeq \frac{1}{4N\sqrt3}\exp{\Big(2\pi\sqrt{\frac{N}{6}}\Big)}\,.
\ee
The number of  microstates of any black hole in the BTZ orbit, recalling \eqref{orbit-invariant-mass}, is given by 
 $p(c\Delta^+)\cdot p(c\Delta^-)$ and the entropy $S$ by
\be\label{microstate-counting}
 S = \ln p(c\Delta^+) + \ln p(c\Delta^-)= 2\pi \Big(\sqrt{\frac{c\Delta^+}{6}} +  \sqrt{\frac{c \Delta^-}{6}}\Big) -\ln (c\Delta^+)-\ln(c\Delta^-)+\cdots 
\ee
where we have assumed $c\Delta^\pm\gg 1$, and $\cdots$ denotes sub-leading terms in $c\Delta^\pm$.

Given the fact that a generic microstate of a black hole in the BTZ orbit is a direct product of a ``microstate part'' and an asymptotic ``macro-state part'' \emph{cf.} \eqref{Microstate-solution}, one readily sees that the microstate part is an orbit invariant quantity and only knows about orbit invariant (asymptotic) charges $J^\pm_0$. 
Therefore, the entropy is also an orbit invariant quantity. This is of course expected recalling discussions of \cite{Sheikh-Jabbari:2016unm} and that entropy is the conserved charge associated with an exact Killing vector field which generates the Killing horizon \cite{Hajian:2015xlp, Sheikh-Jabbari:2016unm}.

\section{Discussions and outlook}\label{section-discussion}

In this work we have repeated the analysis of the ``horizon fluffs'' proposal\footnote{In comparison with the ``fuzzball'' proposal by S. Mathur \cite{fuzzball}, we may call this proposal ``fluff ball''. See discussions of \cite{Afshar:2016uax} for the comparison between the two proposals.} of \cite{Afshar:2016uax} for  generic black holes in the class of Ba\~nados geometries. 
The ``horizon fluffs'' proposal has three main steps:
\begin{itemize}
\item[(1)]\vskip -4mm Identifying the given black hole as a state in the  Hilbert space of {asymptotic} symmetry algebra $\cH_{\cB\cH}$. 
\item[(2)] \vskip -2mm Identifying the near horizon algebra and the Hilbert space of near horizon soft hairs $\HNH$.
\item[(3)] \vskip -2mm Providing asymptotic to near horizon embedding map \eqref{L0-map}, which given the commutation relations in the algebras, induces \eqref{NH-infty-map}.   {This latter in turn provides the relation between the two different presentations of the near horizon algebra in terms of $\bJ$ or $\bcJ$'s.}
 \end{itemize} \vskip -4mm
The last step is the crucial part which enables us to construct all states in $\HNH$ which correspond to the same ``asymptotic black hole state''. These states are what we identified as black hole microstates. 

There are very rigorous constructions and analysis for the steps (1) and (2). The step (3), while still lacking a rigorous proof, is a very well-justified one.
Some early justifications and arguments were already given by Ba\~nados in \cite{Banados-map}. Some more comments are:
\begin{itemize}
\item \vskip -5mm The asymptotic symmetry algebra, as discussed in \cite{Compere:2015knw}, is also symplectic and may hence be defined everywhere in the 3d geometry, including at the horizon. 
\item \vskip -2mm  {Although the near horizon Hilbert space $\HNH$ which is based on $\bcJ_n^\pm$ seems to contain more states than the asymptotic Hilbert space associated with AdS$_3$ black holes $\cH_{\cB\cH}$, $\HNH$ and $\cH_{\cB\cH}$ are both subsets of the Hilbert space of unitary representations (coadjoint orbits) of Brown-Henneaux Virasoro algebra $\cH_{Vir}$. 
We note that $\cH_{Vir}$ includes states which correspond to geometries which are not black holes, and are associated with particles on AdS$_3$ (conic defects). Specifically,
\bi
\item 
\vskip -2mm  $\cH_{\cB\cH}$ includes coadjoint orbits whose value of $J^\pm_0$ can be an arbitrary real number. However, as a result of \eqref{NH-infty-map}, we learn that $\cH_{\cN\cH}$ only contains orbits whose $J^\pm_0$ are ``quantised'', as given in \eqref{orbit-invariant-mass}. This latter, as a gain seen from \eqref{orbit-invariant-mass}, means that our horizon fluffs proposal yields a ``Bohr-type quantisation'' on the mass and angular momentum of the corresponding black hole, specifically, $\Delta^\pm\in\mathbb{Z}/c$.
\item 
\vskip -2mm Moreover, \eqref{NH-infty-map} and the fact that $\HNH$ is the set of near horizon soft hairs (near horizon zero energy states), implies that $\HNH$ includes a specific subclass of elliptic orbits in $\cH_{Vir}$, those whose $\nu$ parameter is quantised as $r/c, r=1,2,\cdots, c-1$.
The representative elements in these orbits (and not their Virasoro excitations) are exactly our near horizon fluffs, microstates of the given AdS$_3$ black hole. 
\item \vskip -2mm If we denote the set of  horizon fluffs by $\cH_\cF$, the above makes it clear that $\cH_\cF\subset \HNH\subset \cH_{Vir}$, and that $\cH_\cF$ and $\cH_{\cB\cH}$ only overlap on the vacuum state.
\ei }
\item \vskip -2mm  {The ``semi-classical Bohr-type'' quantisations are of course confirmed by the specific cases where we know the full quantum theory. That is, for the known cases of the AdS$_3$/CFT$_2$. In the string theory realisations of AdS$_3$/CFT$_2$, e.g. in the D1D5 system \cite{Spectral-flow-1,Spectral-flow-2}, there are usually some restrictions on the spectrum of states appearing in $\cH_{Vir}$, e.g. the conic defects appear in the spectral flow between R and NS vacua, are expected to have $L_0=-6r/c, r=0,1,\cdots, c/6$ \cite{Spectral-flow-2}. (Note that these examples are supersymmetric and the extra factor of 6 compared to our case is due to the fact that we are considering pure {AdS$_3$} gravity case.)}
 \end{itemize}\vskip -5mm
It is of course desirable to make this step (3) as rigorous as the other two steps and  make the relation between the $\HNH$ and $\cH_{Vir}$ more clear. We hope to address this in \cite{Upcoming}.

The analysis of this work, besides extending those of \cite{Afshar:2016uax} (which was made for BTZ black holes) to generic AdS$_3$ black holes, clarifies further the proposal and its technical aspects. In particular it makes the construction of the asymptotic and near horizon Hilbert spaces more explicit. The main result of this work is to identify explicitly    the microstates of black holes in the family of Ba\~nados geometries and establish the fact that the entropy is an ``orbit invariant'' quantity. 
Of course the orbit invariance of entropy and other thermodynamical properties and the first law of thermodynamics for the Ba\~nados family was already stated and argued for  
in the semiclassical analysis \cite{Sheikh-Jabbari:2016unm, Compere:2015knw}  and this work provides a more microscopic description of that.

One of the interesting side results of this work is to show that Ba\~nados geometries are in general coherent states of extended asymptotic symmetry algebra \cite{Afshar:2016wfy}. This is in agreement with the general expectation that geometries, as classical notions, should correspond to coherent states of the underlying quantum system. Here, we of course do not have the full description of pure AdS$_3$ quantum gravity and its dynamics. However, if this theory makes sense (see \cite{Witten-AdS3}), our near horizon Hilbert space $\HNH$ should contain a good part of its Hilbert space. In view of the issues with formulation of AdS$_3$ quantum gravity \cite{Witten-AdS3}, one may wonder if states which are in $\HNH$  but not in $\cH_{\cB\cH}$ may be relevant to the resolution of problems faced there.

 \section*{Acknowledgements}
We would like to thank  Steve Carlip, Finn Larsen and Junbao Wu  for  comments and discussions. We would like to especially thank Hamid Afshar and Daniel Grumiller for many discussions, comments and careful reading of the manuscript.   We also acknowledge the scientific atmosphere of three recent conferences and workshops: \emph{Topics in three dimensional gravity,} March 2016, the Abdus Salam ICTP, Trieste, Italy; \emph{Recent trends in string theory}, May 2016, IPM, Tehran and \emph{Recent developments in symmetries and (super)gravity theories}, June 2016, Bogazici University, Istanbul. The work of  M.M. Sh-J is supported in part by Allameh Tabatabaii Prize Grant of Iran's National Elites Foundation,
the SarAmadan grant of Iranian vice presidency in science and technology and the ICTP network project NET-68.


%


\begin{thebibliography}{99}%

\bibitem{Afshar:2016uax} 
  H.~Afshar, D.~Grumiller and M.~M.~Sheikh-Jabbari,
 ``Near Horizon Soft Hairs as Microstates of Three Dimensional Black Holes,''
  arXiv:1607.00009 [hep-th].

\bibitem{Strominger:1996sh} 
  A.~Strominger and C.~Vafa,
  ``Microscopic origin of the Bekenstein-Hawking entropy,''
  Phys.\ Lett.\ B {\bf 379}, 99 (1996)
  [hep-th/9601029].


\bibitem{Sen:2007qy} 
  A.~Sen,
  ``Black Hole Entropy Function, Attractors and Precision Counting of Microstates,''
  Gen.\ Rel.\ Grav.\  {\bf 40}, 2249 (2008)
  [arXiv:0708.1270 [hep-th]].

\bibitem{BMS} 
  H.~Bondi, M.~G.~J.~van der Burg and A.~W.~K.~Metzner,
  ``Gravitational waves in general relativity. 7. Waves from axisymmetric isolated systems,''
  Proc.\ Roy.\ Soc.\ Lond.\ A {\bf 269}, 21 (1962).

R.~K.~Sachs,
  ``Gravitational waves in general relativity. 8. Waves in asymptotically flat space-times,''
  Proc.\ Roy.\ Soc.\ Lond.\ A {\bf 270}, 103 (1962);  ``Asymptotic symmetries in gravitational theory,''
    Phys.\ Rev.\  {\bf 128}, 2851 (1962).


\bibitem{Brown-Henneaux}
J. D. Brown and M. Henneaux, ``Central Charges in the
Canonical Realization of Asymptotic Symmetries: An
Example from Three-Dimensional Gravity'' Commun.
Math. Phys. 104, 207 (1986).




\bibitem{Lee-Wald} 
  J.~Lee and R.~M.~Wald,
  ``Local symmetries and constraints,''
  J.\ Math.\ Phys.\  {\bf 31}, 725 (1990).

  V.~Iyer and R.~M.~Wald,
    ``Some properties of Noether charge and a proposal for dynamical black hole entropy,''
    Phys.\ Rev.\ D {\bf 50}, 846 (1994)
    [gr-qc/9403028].


\bibitem{Barnich-Brandt} 
  G.~Barnich and F.~Brandt,
 ``Covariant theory of asymptotic symmetries, conservation laws and central charges,''
  Nucl.\ Phys.\ B {\bf 633}, 3 (2002)
  doi:10.1016/S0550-3213(02)00251-1
  [hep-th/0111246].

\bibitem{Thesis-Seraj-Hajian} 
  K.~Hajian,
  ``On Thermodynamics and Phase Space of Near Horizon Extremal Geometries,''
  arXiv:1508.03494 [gr-qc].

  A.~Seraj,
 ``Conserved charges, surface degrees of freedom, and black hole entropy,''
  arXiv:1603.02442 [hep-th].

\bibitem{Sheikh-Jabbari:2016lzm} 
  M.~M.~Sheikh-Jabbari,
  ``Residual Diffeomorphisms and Symplectic Hair on Black Holes,''
  arXiv:1603.07862 [hep-th].
  
  
\bibitem{My-seminar}
M.M. Sheikh-Jabbari, \textit{``Residual Diffeomorphisms and Symplectic Hair on Black Holes''}, seminar presented in workshop on \emph{Recent developments in symmetries and (super)gravity theories}, June 2016,  Bogazici Uni. Istanbul.


  M.~Mirbabayi and M.~Porrati,
  ``Shaving off Black Hole Soft Hair,''
  arXiv:1607.03120 [hep-th].

\bibitem{HPS} 
  S.~W.~Hawking, M.~J.~Perry and A.~Strominger,
  ``Soft Hair on Black Holes,''
  Phys.\ Rev.\ Lett.\  {\bf 116}, no. 23, 231301 (2016)
  [arXiv:1601.00921 [hep-th]].


  G.~Comp\`ere,
  ``Bulk supertranslation memories: a concept reshaping the vacua and black holes of general relativity,''
  arXiv:1606.00377 [hep-th].


\bibitem{no-hair} 
R.~Ruffini and J.~A.~Wheeler,
  ``Introducing the black hole,''
  Phys.\ Today {\bf 24}, no. 1, 30 (1971).


P.~T.~Chrusciel,
 ``'No hair' theorems: Folklore, conjectures, results,''
  Contemp.\ Math.\  {\bf 170}, 23 (1994)
  [gr-qc/9402032].

 J.~D.~Bekenstein,
   ``Black hole hair: 25 - years after,''
   In *Moscow 1996, 2nd International A.D. Sakharov Conference on physics* 216-219
   [gr-qc/9605059].
 




\bibitem{Kerr/CFT} 
  M.~Guica, T.~Hartman, W.~Song and A.~Strominger,
  ``The Kerr/CFT Correspondence,''
  Phys.\ Rev.\ D {\bf 80}, 124008 (2009)
  [arXiv:0809.4266 [hep-th]].


G.~Comp\`ere,
 ``The Kerr/CFT correspondence and its extensions: a comprehensive review,''
  Living Rev.\ Rel.\  {\bf 15}, 11 (2012)
  [arXiv:1203.3561 [hep-th]].

\bibitem{BMS-algebra} 
 
 
  G.~Barnich and C.~Troessaert,
  ``Symmetries of asymptotically flat 4 dimensional spacetimes at null infinity revisited,''
  Phys.\ Rev.\ Lett.\  {\bf 105}, 111103 (2010)
  [arXiv:0909.2617 [gr-qc]];
 ``BMS charge algebra,''
  JHEP {\bf 1112}, 105 (2011)
  [arXiv:1106.0213 [hep-th]].

G.~Barnich, A.~Gomberoff and H.~A.~Gonzalez,
  ``The Flat limit of three dimensional asymptotically anti-de Sitter spacetimes,''
  Phys.\ Rev.\ D {\bf 86} (2012) 024020
  [arXiv:1204.3288 [gr-qc]].


  E.~E.~Flanagan and D.~A.~Nichols,
  ``Conserved charges of the extended Bondi-Metzner-Sachs algebra,''
  arXiv:1510.03386 [hep-th].

\bibitem{BMS-Orbits} 
 P.~J.~McCarthy and E.~Melas,
  ``On irreducible representations of the ultrahyperbolic BMS group,''
  Nucl.\ Phys.\ B {\bf 653}, 369 (2003).

 G.~Barnich and B.~Oblak,
 ``Notes on the BMS group in three dimensions: I. Induced representations,''
  JHEP {\bf 1406}, 129 (2014)
  [arXiv:1403.5803 [hep-th];
  ``Notes on the BMS group in three dimensions: II. Coadjoint representation,''
  JHEP {\bf 1503}, 033 (2015)
  [arXiv:1502.00010 [hep-th]].

 A.~Garbarz and M.~Leston,
  ``Quantization of BMS$_3$ orbits: a perturbative approach,''
  Nucl.\ Phys.\ B {\bf 906}, 133 (2016)
  [arXiv:1507.00339 [hep-th]].

\bibitem{Garbarz} 
  A.~Garbarz and M.~Leston,
 ``Classification of Boundary Gravitons in AdS$_3$ Gravity,''
  JHEP {\bf 1405}, 141 (2014), [arXiv:1403.3367 [hep-th]].  

  G.~Barnich and B.~Oblak,
  ``Holographic positive energy theorems in three-dimensional gravity,''
  Class.\ Quant.\ Grav.\  {\bf 31}, 152001 (2014)
  [arXiv:1403.3835 [hep-th]].

\bibitem{Carlip-NH-entropy} 
  S.~Carlip,
  ``Near horizon conformal symmetry and black hole entropy,''
  Phys.\ Rev.\ Lett.\  {\bf 88}, 241301 (2002)
  [gr-qc/0203001].


\bibitem{NH-algebra-1} 
L.~Donnay, G.~Giribet, H.~A.~Gonzalez and M.~Pino,
 ``Supertranslations and Superrotations at the Black Hole Horizon,''
  Phys.\ Rev.\ Lett.\  {\bf 116}, no. 9, 091101 (2016)
  doi:10.1103/PhysRevLett.116.091101
  [arXiv:1511.08687 [hep-th]];
  ``Extended Symmetries at the Black Hole Horizon,''
  arXiv:1607.05703 [hep-th].



\bibitem{NH-algebra-2} 
H.~Afshar, S.~Detournay, D.~Grumiller and B.~Oblak,
  ``Near-Horizon Geometry and Warped Conformal Symmetry,''
  JHEP {\bf 1603}, 187 (2016)
  [arXiv:1512.08233 [hep-th]].
\bibitem{NH-algebra-3}

  M.~R.~Setare and H.~Adami,
 ``The Heisenberg algebra as near horizon symmetry of the black flower solutions of Chern-Simons-like theories of gravity,''
  arXiv:1606.05260 [hep-th]; ``Lorentz-diffeomorphism quasi-local conserved charges and Virasoro algebra in Chern–Simons-like theories of gravity,''
    Nucl.\ Phys.\ B {\bf 909}, 345 (2016)
    [arXiv:1511.01070 [hep-th]].

  M.~Hotta, J.~Trevison and K.~Yamaguchi,
  ``Gravitational Memory Charges of Supertranslation and Superrotation on Rindler Horizons,''
  arXiv:1606.02443 [gr-qc].


 D.~Grumiller, A.~Perez, S.~Prohazka, D.~Tempo and R.~Troncoso,
  ``Higher Spin Black Holes with Soft Hair,''
  arXiv:1607.05360 [hep-th].



\bibitem{Afshar:2016wfy} 
  H.~Afshar, S.~Detournay, D.~Grumiller, W.~Merbis, A.~Perez, D.~Tempo and R.~Troncoso,
  ``Soft Heisenberg hair on black holes in three dimensions,''
  Phys.\ Rev.\ D {\bf 93}, no. 10, 101503 (2016)
 [arXiv:1603.04824 [hep-th]].

\bibitem{Lucietti}

 H.~K.~Kunduri, J.~Lucietti and H.~S.~Reall,
 ``Near-horizon symmetries of extremal black holes,''
  Class.\ Quant.\ Grav.\  {\bf 24}, 4169 (2007)
  [arXiv:0705.4214 [hep-th]].

 H.~K.~Kunduri and J.~Lucietti,
 ``Classification of near-horizon geometries of extremal black holes,''
  Living Rev.\ Rel.\  {\bf 16}, 8 (2013)
  [arXiv:1306.2517 [hep-th]].


\bibitem{Carlip:2004ba} 
  S.~Carlip,
  Living Rev.\ Rel.\  {\bf 8}, 1 (2005)
  doi:10.12942/lrr-2005-1
  [gr-qc/0409039].

\bibitem{BTZ}
M.~Ba\~nados, C.~Teitelboim and J.~Zanelli, ``The Black hole in three-dimensional space-time,''
  Phys.\ Rev.\ Lett.\  {\bf 69}, 1849 (1992),
  [hep-th/9204099].

M.~Ba\~nados, M.~Henneaux, C.~Teitelboim and J.~Zanelli,
  ``Geometry of the (2+1) black hole,''
  Phys.\ Rev.\ D {\bf 48}, 1506 (1993)
  [gr-qc/9302012].



\bibitem{Banados}
M.~{Ba{\~n}ados}, ``{Three-Dimensional Quantum Geometry and Black Holes},''
\href{http://www.arXiv.org/abs/hep-th/9901148}{{\tt hep-th/9901148}}.

\bibitem{Sheikh-Jabbari:2014nya} 
  M.~M.~Sheikh-Jabbari and H.~Yavartanoo,
  ``On quantization of AdS$_{3}$ gravity I: semi-classical analysis,''
  JHEP {\bf 1407}, 104 (2014)
  [arXiv:1404.4472 [hep-th]].


\bibitem{Sheikh-Jabbari:2016unm} 
  M.~M.~Sheikh-Jabbari and H.~Yavartanoo,
  ``On 3d Bulk Geometry of Virasoro Coadjoint Orbits: Orbit invariant charges and Virasoro hair on locally AdS3 geometries,''
  arXiv:1603.05272 [hep-th].

\bibitem{Compere:2015knw} 
  G.~Comp\`ere, P.~J.~Mao, A.~Seraj and M.~M.~Sheikh-Jabbari,
  ``Symplectic and Killing symmetries of AdS$_{3}$ gravity: holographic vs boundary gravitons,''
  JHEP {\bf 1601}, 080 (2016)
  [arXiv:1511.06079 [hep-th]].



\bibitem{Vir-Orbits}
A.~A.~Kirillov, Funct. Anal. Appl. 15 (2) (1981) 135; Springer Lecture Notes in
Mathematics, vol. 970 (1982) 101.

  G.~Segal,
  ``Unitarity Representations of Some Infinite Dimensional Groups,''
  Commun.\ Math.\ Phys.\  {\bf 80}, 301 (1981).
 
  E.~Witten,
 ``Coadjoint Orbits of the Virasoro Group,''
  Commun.\ Math.\ Phys.\  {\bf 114}, 1 (1988).
  doi:10.1007/BF01218287

\bibitem{Balog} 
  J.~Balog, L.~Feher and L.~Palla,
  ``Coadjoint orbits of the Virasoro algebra and the global Liouville equation,''
  Int.\ J.\ Mod.\ Phys.\ A {\bf 13}, 315 (1998)
  doi:10.1142/S0217751X98000147
  [hep-th/9703045].

\bibitem{Cedric} 
 G.~Barnich and C.~Troessaert,
  ``Aspects of the BMS/CFT correspondence,''
  JHEP {\bf 1005}, 062 (2010)
  [arXiv:1001.1541 [hep-th]].


  C.~Troessaert,
 ``Enhanced asymptotic symmetry algebra of $AdS$$_{3}$,''
  JHEP {\bf 1308}, 044 (2013)
  [arXiv:1303.3296 [hep-th]].


\bibitem{Daniel}
D.~Grumiller and M.~Riegler,
  ``Most general AdS$_3$ boundary conditions,''
  arXiv:1608.01308 [hep-th].


\bibitem{Hill-Eq}
W. Magnus and S. Winkler,  \textit{Hill's Equation}, (2004), Dover publications.



\bibitem{Hajian:2015xlp} 
  K.~Hajian and M.~M.~Sheikh-Jabbari,
  ``Solution Phase Space and Conserved Charges: A General Formulation for Charges Associated with Exact Symmetries,''
  Phys.\ Rev.\ D {\bf 93}, no. 4, 044074 (2016)
  [arXiv:1512.05584 [hep-th]].


\bibitem{Sheikh-Jabbari:2016znt} 
  M.~M.~Sheikh-Jabbari and H.~Yavartanoo,
  ``Excitation Entanglement Entropy in 2d Conformal Field Theories,''
  arXiv:1605.00341 [hep-th].

\bibitem{Upcoming}
H.R. Afshar, D. Grumiller, F. Larsen, M.M. Sheikh-Jabbari, H. Yavartanoo, ``{More on Horizon Fluffs Proposal: Near Horizon vs Asypmtotic Hilbert Spaces},'' \emph{To appear.} 

\bibitem{Song-Compere-Strominger} 
  G.~Comp\`ere, W.~Song and A.~Strominger,
  ``New Boundary Conditions for AdS3,''
  JHEP {\bf 1305}, 152 (2013)
  [arXiv:1303.2662 [hep-th]].

\bibitem{ASS-work-in-progress}

H. Afshar, A. Seraj, M.M. Sheikh-Jabbari,  \emph{Work in progress}. 


\bibitem{symplectic} 
 G.~Comp\`ere, M.~Guica and M.~J.~Rodriguez,
  ``Two Virasoro symmetries in stringy warped AdS$_{3}$,''
  JHEP {\bf 1412}, 012 (2014), [arXiv:1407.7871 [hep-th]].

 G.~Comp\`ere, K.~Hajian, A.~Seraj and M.~M.~Sheikh-Jabbari,
  ``Extremal Rotating Black Holes in the Near-Horizon Limit: Phase Space and Symmetry Algebra,''
  Phys.\ Lett.\ B {\bf 749}, 443 (2015), [arXiv:1503.07861 [hep-th]];
Wiggling Throat of Extremal Black Holes,''
  JHEP {\bf 1510}, 093 (2015),  [arXiv:1506.07181 [hep-th]].

\bibitem{Mitchell} 
  J.~M.~Mitchell,
  ``Where are the BTZ Black Hole Degrees of Freedom? The Rotating Case,''
  Class.\ Quant.\ Grav.\  {\bf 33}, no. 13, 135003 (2016)
  [arXiv:1510.01033 [gr-qc]].


\bibitem{Banados-94} 
  M.~Ba\~nados,
  ``Global charges in Chern-Simons field theory and the (2+1) black hole,''
  Phys.\ Rev.\ D {\bf 52}, 5816 (1996)
  [hep-th/9405171].

\bibitem{Banados:1998ta} 
  M.~Ba\~nados, T.~Brotz and M.~E.~Ortiz,
  ``Boundary dynamics and the statistical mechanics of the (2+1)-dimensional black hole,''
  Nucl.\ Phys.\ B {\bf 545}, 340 (1999)
  [hep-th/9802076].

\bibitem{Afshar:2014rwa} 
  H.~Afshar, A.~Bagchi, S.~Detournay, D.~Grumiller, S.~Prohazka and M.~Riegler,
  ``Holographic Chern-Simons Theories,''
  Lect.\ Notes Phys.\  {\bf 892}, 311 (2015)
  [arXiv:1404.1919 [hep-th]].



\bibitem{Banados-map} 
  M.~Ba\~nados,
 ``Embeddings of the Virasoro algebra and black hole entropy,''
  Phys.\ Rev.\ Lett.\  {\bf 82}, 2030 (1999)
  [hep-th/9811162].


\bibitem{Witten-AdS3} 
  E.~Witten,
  ``Three-Dimensional Gravity Revisited,''
  arXiv:0706.3359 [hep-th].
  
 A.~Maloney and E.~Witten,
  ``Quantum Gravity Partition Functions in Three Dimensions,''
  JHEP {\bf 1002}, 029 (2010)
  [arXiv:0712.0155 [hep-th]].

For further discussions on this point see the recent paper:\\
J.~B.~Bae, K.~Lee and S.~Lee,``Bootstrapping Pure Quantum Gravity in AdS3,''
  arXiv:1610.05814 [hep-th].

\bibitem{Spectral-flow-1} 
J.~R.~David, G.~Mandal, S.~Vaidya and S.~R.~Wadia,
``Point mass geometries, spectral flow and AdS(3) - CFT(2) correspondence,''
  Nucl.\ Phys.\ B {\bf 564}, 128 (2000)
  [hep-th/9906112].


J.~R.~David, G.~Mandal and S.~R.~Wadia,
 ``Microscopic formulation of black holes in string theory,''
  Phys.\ Rept.\  {\bf 369}, 549 (2002)
  [hep-th/0203048].

\bibitem{Spectral-flow-2} 
J.~M.~Maldacena and L.~Maoz,
 ``Desingularization by rotation,''
  JHEP {\bf 0212}, 055 (2002)
  [hep-th/0012025].

  O.~Lunin, J.~M.~Maldacena and L.~Maoz,
  ``Gravity solutions for the D1-D5 system with angular momentum,''
  hep-th/0212210.


\bibitem{Carlip:1998qw} 
  S.~Carlip,
  ``What we don't know about BTZ black hole entropy,''
  Class.\ Quant.\ Grav.\  {\bf 15}, 3609 (1998)
  [hep-th/9806026].


\bibitem{fuzzball}
S.~D.~Mathur,
  ``Fuzzballs and the information paradox: A Summary and conjectures,''
  arXiv:0810.4525 [hep-th]; ``The Fuzzball proposal for black holes: An Elementary review,''
    Fortsch.\ Phys.\  {\bf 53}, 793 (2005)
    [hep-th/0502050].



\end{thebibliography}
\end{document}